\documentclass[sigconf]{acmart}
\AtBeginDocument{%
  }
\usepackage{multirow} 
\usepackage{xcolor}
\usepackage[most]{tcolorbox}
\usepackage{tabularx}
\usepackage{stfloats}
\usepackage{booktabs}
\usepackage{array}
\usepackage{graphicx}
\newcolumntype{L}{>{\raggedright\arraybackslash}X} 
\newcolumntype{C}{>{\centering\arraybackslash}X}  

\setcopyright{acmlicensed}
\copyrightyear{2026}
\acmYear{2026}
\acmDOI{XXXXXXX.XXXXXXX}

\acmConference[DIS '26]{Designing Interactive Systems Conference 2026}{June 13--17, 2026}{Singapore}

\acmISBN{978-1-4503-XXXX-X/2018/06}

\begin{document}


\title{Reshaping Inclusive Interpersonal Dynamics through Smart Glasses in Mixed-Vision Social Activities}


\author{Jieqiong Ding}
\authornote{These authors contributed equally to this work.}
\affiliation{%
  \institution{The Future Laboratory, Tsinghua University}
  \city{Beijing}
  \country{China}
}
\affiliation{%
  \institution{Human Centered Design \& Engineering, University of Washington}
  \city{Seattle}
  \state{Washington}
  \country{United States}
}

\author{Yumo Zhang}
\authornotemark[1]
\affiliation{%
  \institution{The Future Laboratory, Tsinghua University}
  \city{Beijing}
  \country{China}
}
\affiliation{%
  \institution{School of Design, Hong Kong Polytechnic University}
  \city{Hong Kong}
  \country{China}
}

\author{Xiuqi Tommy Zhu}
\affiliation{%
  \institution{College of Arts, Media and Design, Northeastern University}
  \city{Boston}
  \state{Massachusetts}
  \country{United States}
}

\author{Kaige Yang}
\affiliation{%
  \institution{The Future Laboratory, Tsinghua University}
  \city{Beijing}
  \country{China}
}
\affiliation{%
  \institution{Central Academy of Fine Arts}
  \city{Beijing}
  \country{China}
}

\author{Yuqing Wei}
\affiliation{%
  \institution{The Future Laboratory, Tsinghua University}
  \city{Beijing}
  \country{China}
}
\affiliation{%
  \institution{School of Design, Shanghai Jiao Tong University}
  \city{Shanghai}
  \country{China}
}

\author{Shiyi Wang}
\affiliation{%
  \institution{Academy of Arts \& Design, Tsinghua University}
  \city{Beijing}
  \country{China}
}

\author{Yishan Liu}
\affiliation{%
  \institution{Academy of Arts \& Design, Tsinghua University}
  \state{Beijing}
  \country{China}
}

\author{Yang Jiao}
\email{jiaoyang7@tsinghua.edu.cn}
\affiliation{%
  \institution{The Future Laboratory, Tsinghua University}
  \city{Beijing}
  \country{China}
}
\authornote{Corresponding author.}

\renewcommand{\shortauthors}{Ding et al.}

\begin{abstract}

Meaningful social interaction is vital to well-being, yet Blind and Low Vision (BLV) individuals face persistent barriers when collaborating with sighted peers due to inaccessible visual cues. While most wearable assistive technologies emphasize individual tasks, smart glasses introduce opportunities for real-time, contextual support in social settings. To explore how smart glasses affect the interpersonal dynamics and support inclusion in mixed-vision groups, we developed a smart glasses–based system \textit{CollabLens} as a technology probe, and employed it in four workshop sessions. We found that smart glasses can meaningfully support inclusive collaboration and provide users with greater flexibility and control. However, while sighted participants viewed smart glasses as a medium reshaping how assistance was negotiated with great potential, BLV participants primarily valued smart glasses for independent task completion compared to social purposes. We concluded by discussing and synthesizing challenges and opportunities for designing smart glasses that foster natural social dynamics in future inclusive settings.
\end{abstract}



\begin{teaserfigure}
  \includegraphics[width=\textwidth]{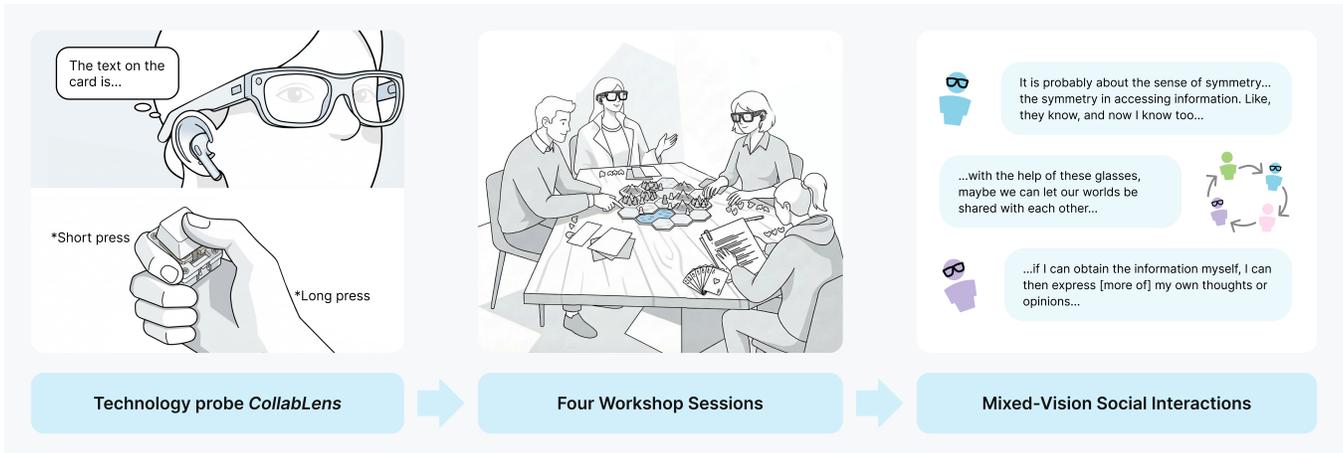}
  \caption{We explored how smart glasses reshape interpersonal dynamics and for inclusive participation in mixed-vision activities. Through deploying our technology probe \textit{CollabLens} in a workshop study, we found that smart glasses enhanced autonomy for BLV participants and reduced the assistance burden on sighted peers, facilitating a more reciprocal dynamic. We highlighted the importance of evolving toward flexible interdependence for assistive technologies.}
  \Description{We explored how smart glasses reshape interpersonal dynamics and for inclusive participation in mixed-vision activities. Through deploying our technology probe CollabLens in a workshop study, we found that smart glasses enhanced autonomy for BLV participants and reduced the assistance burden on sighted peers, facilitating a more reciprocal dynamic. We highlighted the importance of evolving toward flexible interdependence for assistive technologies.}
  \label{fig:teaser}
\end{teaserfigure}

\received{20 February 2007}
\received[revised]{12 March 2009}
\received[accepted]{5 June 2009}

\maketitle
\section{Introduction}
Participating in social activities with sighted peers plays a crucial role in supporting blind or low-vision (BLV) individuals by enhancing communication, emotional bonds, and overall participation in cultural and community life \cite{shah2020association, brunes2019loneliness, takesue2021social}.
However, BLV individuals encounter additional barriers during these mixed-vision social activities, including accessing, interpreting, and coordinating information with visual cues, which can disrupt their experiences or limit opportunities for engagement \cite{fryer2021accessing, guerreiro2013blind, bolesnikov2022understanding, klauke2023impact}. These challenges eventually lead to an environment where BLV participants experience reduced feelings of \textbf{inclusion}, discouraging participation in mixed-vision activities \cite{zhu_understanding_nodate, metatla_voice_2019, piedade_inclusion_2024}.


HCI researchers have explored assistive technologies (AT) such as tactile interfaces and screen readers \cite{raynal2024flexiboard, jung2024accessible} to support mixed-vision interaction between BLV individuals and sighted peers across various scenarios, such as learning, coding, and writing \cite{das_it_2019, das_co11ab_2022, potluri_codewalk_2022, kane2015oneview, metatla_voice_2019, piedade_inclusion_2024, branham_collaborative_2015}.
However, most of them focused on structured or task-oriented scenarios. Although some studies paid attention to the social and entertainment needs of BLV people by designing inclusive mixed-vision activities \cite{cheiran2011inclusive, Collins_2024, McElligott_2004, Sánchez_Lumbreras_Cernuzzi_2001, Sánchez_Sáenz_Ripoll_2009}, such as building mixed-vision social game to encourage collaboration of BLV and sighted players \cite{Xieraili_Wang_Guo_Liu_Qiu_2025, Reinhardt_Silveira_Tait_Loke_Jones_Holloway_2023, Luo_Liu_Hu_2024}, their contributions are limited to single application scenario rather than ubiquitous interaction model.

Powered by Large Multimodal Models (LMM), AT are now capable of providing real-time visual interpretation through voice feedback that adapts to various scenarios \cite{chang_probing_2025, xie_beyond_2025, chang2024, ning2025aroma}.
Compared to traditional hand-held AT such as smartphones, smart glasses \footnote{We use the term smart glasses to refer to voice-only interaction glasses with multimodal input ability, without augmented reality displays.}
provide unique opportunities for real-time social interaction through hands-free affordances \cite{lee_interaction_2018, waisberg_meta_2024, chen_visimark_2025}. 
While recent research has demonstrated smart glasses' potential to support BLV users in individual scenarios \cite{udayakumar_artificial_2025, varshney_evaluating_nodate, li_scoping_2022, gamage_smart_2025}, few studies have explored their potential in addressing the complex, dynamic challenges of mixed-vision social activities, which extend beyond technical accuracy to include providing context-aware and timely support \cite{tang_everyday_2025}. 
This motivates us to further explore the potential of this emerging technology in mixed-vision social activities by understanding its current practices through the following three research questions.

\begin{itemize}
    

\item RQ1: How do BLV individuals engage with smart glasses during mixed-vision social activities?
\item RQ2: What challenges and opportunities emerge when using smart glasses to support mixed-vision social activities?
\item RQ3: How do BLV individuals and sighted peers perceive the role of smart glasses in shaping inclusive mixed-vision social experiences?

\end{itemize}

To answer both questions, we first developed \textit{CollabLens}, an LMM-based voice assistant designed to interpret visual information in real time through an intuitive interaction interface, serving as a technology probe on smart glasses. We then deployed it in four mixed-vision workshop sessions where BLV and sighted participants join in tabletop activities collaboratively through mixed-method analysis.
We found that BLV participants positively viewed smart glasses as a convenient and hands-free addition to their existing support network, enabling access to visually centered information. They adopted smart glasses primarily to accomplish tasks on their own to relieved pressure of relying on sighted peers. Nonetheless, they held reserved perspectives on whether smart glasses can sufficiently support inclusive experiences in everyday life, revealing the importance to design smart glasses within their existing assistive strategies. Comparatively, sighted participants viewed smart glasses as a promising tool that facilitates social inclusion and enhances interpersonal connections with their BLV peers. They also mentioned challenges in their adaptation to the introduction of the new technology. We also identified challenges of integrated smart glasses to seamless social interaction, including system reliability, social embarrassment, and long-term wearing comfort. We then discussed how such devices enrich the current assistive network beyond independence and reciprocal inclusion dynamics, followed by proposing the design implications for future LMM-based smart glasses and AT that aim to support inclusive social collaboration.

Our research contributes to the growing understanding of social accessibility by: (1) presenting \textit{CollabLens}, a technology probe implemented on smart glasses with an LMM-based voice interface; (2) offering empirical insights into how BLV individuals and sighted peers use and perceive smart glasses during mixed-vision social activities; and (3) articulating design opportunities and challenges for developing smart glasses that foster inclusive and natural social interactions across mixed-vision groups.

\section{Related Works}
\subsection{Supporting Mixed Vision Social Interactions}
Mixed-vision social activities are vital for blind and low-vision (BLV) individuals, as they foster communication, emotional connection, and broader inclusion in cultural and community interaction \cite{shah2020association, brunes2019loneliness}. Thus, researchers are calling to promote BLV individuals towards inclusive mixed-vision activities for engaging in social experiences \cite{takesue2021social, zhu_understanding_nodate, ran_understanding_2025}. However, BLV individuals face a number of challenges in these mixed-vision social activities. Particularly, BLV individuals often struggle with information gaps such as accessing (e.g., difficulty noticing or reading printed notices about events \cite{fryer2021accessing}), outputting (e.g., challenges in expressing non-visual cues in social media or games \cite{guerreiro2013blind}), and aligning information (e.g., difficulty interpreting others’ intentions or emotional signals \cite{klauke2023impact}).
For example, during collaborative game sessions, BLV participants may need to access task instructions presented visually (e.g., on printed cards or slides) and express their actions or intentions without visual cues (e.g., pointing or gesturing). This leads to activity breakdowns that exclude them from group \cite{bolesnikov2022understanding}. Additionally, the absence of visual cues, such as facial expressions and body language, has been proven to reduce engagement and emotional connection, leading to missing mutual understandings \cite{klauke2023impact}. This is also a challenge in mixed-vision activities to create a sense of group inclusion \cite{zhu_understanding_nodate, metatla_voice_2019, piedade_inclusion_2024}. 


Previous studies have unpacked how BLV individuals increasingly gain opportunities to engage more independently in social participation and collaborative settings through accessible technologies (e.g, tactile interfaces \cite{raynal2024flexiboard} and screen readers \cite{jung2024accessible}) rather than relying on human assistance \cite{mulloy2014assistive}, including such as writing \cite{das_it_2019, das_co11ab_2022}, coding \cite{potluri_codewalk_2022}, learning \cite{kane2015oneview, metatla_voice_2019, piedade_inclusion_2024}, and daily living \cite{branham_collaborative_2015}.
For example, \citet{das_co11ab_2022} built \textit{Co11ab}, a Google Docs extension that provides configurable audio cues to help users identify who is editing and where in a shared document, thereby addressing the information gap in collaborative tasks.
\citet{teng_help_2024} introduces \textit{Help Supporters}, assistive systems that mediate face-to-face collaboration between BLV individuals and sighted strangers by reducing social barriers during help initiation and providing in-situ guidance to improve collaborative communication. However, most of these studies have focused on structured or context-oriented collaboration, where the primary goal is functional coordination rather than shared social experience.

Building upon prior efforts to design inclusive social activities for BLV individuals \cite{cheiran2011inclusive, Xieraili_Wang_Guo_Liu_Qiu_2025, Reinhardt_Silveira_Tait_Loke_Jones_Holloway_2023, Luo_Liu_Hu_2024, cassidy_cuddling_2024}, our work contributes to a more ubiquitous interaction mode that fosters inclusive interpersonal experiences between BLV and sighted users through smart glasses as an adaptive solution across diverse social scenarios.

\subsection{AI-Enhanced Assistive Technologies}

Recent advances in Machine Learning (ML) and Natural Language Processing (NLP) illustrate a growing trend toward Artificial Intelligence (AI) systems capable of visual interpretation and adaptive feedback, as seen in systems like SeeingAI \cite{seeingai2025}, Be My AI \cite{bemyai2025}, and Orcam \cite{orcam2025}. As BLV users rely heavily on voice input and output to access, navigate, and control information \cite{jung2024accessible, metatla_voice_2019, singh2012blind}, traditional voice assistants (VAs) such as Siri and Alexa fail to provide contextual understanding and flexibility in information presentation \cite{abdolrahmani_siri_2018, sayago_apple_2021}, leading to breakdowns that disrupts BLV user flow and diminish their confidence in sustaining engagement with such technologies \cite{brady_investigating_2013}. Large Multimodal Model (LMM) now can tackle these challenges by allowing users to engage in open-ended dialogue, interpret context, and provide human-like responses \cite{mahmood_voice_2025, brown2020language, mahmood2025user}. For BLV individuals, this evolution marks a critical transition from merely functional auditory access to collaborative visual understanding in real-world contexts \cite{xie_beyond_2025, chang_probing_2025, adnin_i_2024, kuzdeuov_chatgpt_2024, tang_everyday_2025}.

Recent research focuses on understanding how LMM enables BLV users to independently obtain visual information in daily life, such as cooking \cite{li_oscar_2025, ning2025aroma}, navigation \cite{zhang_enhancing_2025, xia_ibgs_2022}, and applying makeup \cite{li_more_2025}. For instance, \citet{chang2024} introduced \textit{WorldScribe}, which uses LMM to generate dynamic scene descriptions that pause appropriately during conversations, adapting to both visual and audio contexts in everyday activities on a mobile phone. 

However, only a few works aim to apply AI in mixed-vision activities. For example, \citet{chheda-kothary_engaging_2024} examined how BLV family members engage with their sighted children’s artwork, revealing that emotional bonding is prioritized over visual accuracy and that AI-generated descriptions can support, but not replace, child-led storytelling and multisensory engagement in their family activities. However, it remains unexplored how LMM could affect social interactions between BLV and sighted peers in real time. Our work contributes to this gap by investigating how LMM can support real-time, dynamic mixed-vision interactions as an always-on helper, extending beyond previously studied individual-use scenarios.

\subsection{Smart Glasses for BLV Users}
Powered by LMM, smart glasses can now provide unique opportunities for dynamic social contexts. Unlike traditional handheld tools, smart glasses provide a wearable, hands-free, and flexible platform that better supports natural, everyday activities \cite{lee_interaction_2018, waisberg_meta_2024, chen_visimark_2025}. When specifically focusing on the low vision community, researchers adopt augmented reality (AR) displays on smart glasses. For example, \citet{chen_visimark_2025} developed VisiMark, an AR-based interface on smart glasses that supports landmark perception for people with low vision by providing both spatial overviews and in-situ landmark augmentations. Similarly, \citet{lee_towards_2024} introduced ARSports, a wearable AR research prototype that overlays instance segmentation masks in near real time to improve visual tracking for people with low vision in sports accessibility. By contrast,
\citet{lee_lab_2022} demonstrated that using smart glasses in remote studies with blind participants enables substantially better first-person access to users’ interactions than stationary laptop cameras, while also revealing practical challenges around camera field of view, accessibility support, battery life, and study logistics. However, similarly, recent research on smart glasses primarily explored their potential in individual scenarios \cite{udayakumar_artificial_2025, varshney_evaluating_nodate, li_scoping_2022, gamage_smart_2025}. It remains underexplored how such socially assistive devices are used in everyday social contexts involving other people.

In mixed-vision social activities, the challenge extends beyond technical accuracy to providing context-aware and timely support that balances autonomy and coordination—something current assistive technologies often struggle to achieve \cite{lodewyk_wearables_2025, metatla_voice_2019, cassidy_cuddling_2024}. It is therefore important to understand how BLV individuals adapt to these promising and potential devices. To address this gap, our work empirically examines the performance of smart glasses in mixed-vision social activities, exploring how BLV and sighted participants use and perceive them in such contexts. 

\begin{figure*}[t]
    \centering
    \includegraphics[width=\linewidth]{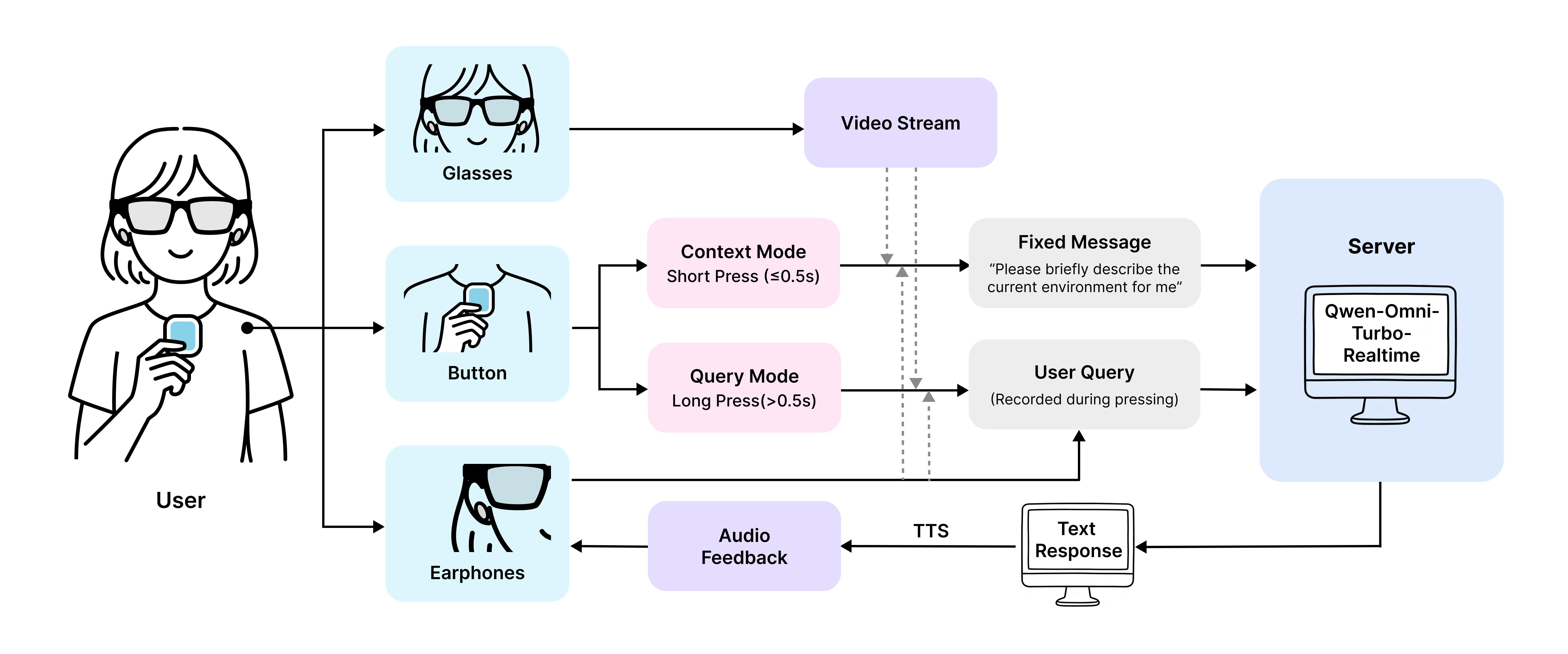}
    \caption{System Pipeline of \textit{CollabLens}. When a request is initiated, the video and audio stream captured will be passed to a local server together with the audio prompts. The user activates the Context Mode or Query Mode by performing short or long press on the button respectively. In Context Mode, a pre-recorded audio with fixed message for environmental description will be serve as the prompt. In Query Mode, the user’s query will be recorded and sent. The input will be processed by Qwen-Omni-Turbo-Realtime, which returns a text response. This text is finally converted to audio and delivered to the user through earphones.}
    \Description{The figure illustrates the System Pipeline of CollabLens. When a request is initiated, the video and audio stream captured will be passed to a local server together with the audio prompts. The user activates the Context Mode or Query Mode by performing short or long press on the button respectively. In Context Mode, a pre-recorded audio with fixed message for environmental description will be serve as the prompt. In Query Mode, the user’s query will be recorded and sent. The input will be processed by Qwen-Omni-Turbo-Realtime, which returns a text response. This text is finally converted to audio and delivered to the user through earphones.}
    \label{fig:implementation}
\end{figure*}
\section{\textit{CollabLens}}


\subsection{Overview of Technology Probe}
\label{why probe}
Since our current goal was to explore and define directions for smart glasses in mixed-vision social interactions, we aimed at creating a technology probe that could flexibly and effectively elicit user needs and interaction patterns in real interpersonal contexts rather than directly building an optimized technical solution. 

A technology probe is a functional prototype of emerging technologies deployed to provoke users' thoughts and inspire understanding of what the technology might be \cite{10.1145/642611.642616}. 
Prior HCI work has demonstrated the feasibility of technology probe studies in identifying opportunities for emerging technologies without constraining participants to predetermined interaction models \cite{Du_2024, Feng_2025}. 
Meanwhile, as our research focuses on understanding users' behaviors and needs and inspiring future work compared to building systems that solve specific problems or excel against baselines, technology probes' exploratory nature align well with our goals. 
Therefore, we followed this approach and developed a smart glasses-based system \textit{CollabLens} to support mixed-vision social interactions.

\subsection{System Design}
\label{system design}

\textit{CollabLens} was first inspired by learning from the successful use of LMM-integrated smart glasses and then gathered design rationale from BLV challenges and needs, particularly in how LMM-integrated assistive technologies support mixed-vision activities.
As we summarized in Section 2, LMM-integrated smart glasses shows advantage of interpreting seamless visual information with natural communication, and could potentially address the crucial information gaps during mixed-vision social activities. With this being said, our CollabLens design followed three design rationales: \textit{(DR1) Keep the interaction simple:} the system should be easy to interact with and representative of current smart glasses to meet the goal of a technology probe. \textit{(DR2) Balancing accessibility and agency:} the system should leverage the agency of BLV individuals instead of only pursuing technical performance that does not respect accessibility considerations. \textit{(DR3) Dynamic context awareness during collaboration:} the system should hold the capability to adapt to the dynamic nature of mixed-vision collaboration and address different challenges in real time. These rationales did not directly inform our subsequent interaction design, but collectively provide a holistic perspective to ensure that \textit{CollabLens} holds the capability to meet the needs of a technology probe for the following user study.


Building on that, we decided to keep voice interaction as the primary vehicle to mimic general commercial smart glasses, while adding two specific interaction scenarios for accessibility considerations: \textbf{1) Context Mode}, where CollabLens delivers concise scene descriptions, and \textbf{2) Query Mode}, where users can ask specific questions to acquire more complex information based on their needs. We also adopted a button-based interaction to provide a tangible interaction mechanism for BLV users. A short press triggers the \textit{Context} Mode without requiring voice input, while a long press triggers the \textit{Query} Mode, recording the user’s input during the press. We aim to create a simple yet familiar interaction that blends smart glasses naturally into the interaction flow rather than drawing attention through overt or intrusive interaction modalities.



\subsection{Implementation}

To achieve real-time environmental perception and sensory augmentation, we implemented \textit{CollabLens} on the RayNeo X2\footnote{Rayneo. \textit{Rayneo Official Site.} Retrieved October 8, 2025 from https://www.rayneo.com.} smart glasses platform. The system pipeline is illustrated in Figure~\ref{fig:implementation}.

To begin with, smart glasses first capture first-person video using a built-in camera. Each frame is downsampled to 320×240 resolution and compressed to 60\% JPEG quality to balance network latency and image fidelity. Frames are streamed at 2 FPS via the Transmission Control Protocol to a local Python server.
Simultaneously, audio is recorded at 24 kHz using the PyAudio library for \cite{pyaudio-doc} low-latency voice processing. 
Both video and audio streams and audio of prompts are transmitted via WebSocket to the Qwen-Omni-Turbo-Realtime API~\cite{qwen-realtime-api}, which we selected as it is an open-source LMM with real-time video understanding capabilities available for our study location. In \textit{Context} Mode, where the user presses the button for less than 0.5 seconds, the system passes a pre-recorded audio input (Please briefly describe the current environment for me). We also provide an additional physical button component that, in \textit{Query} Mode, when the user presses the button for more than 0.5 seconds, their voice query will be recorded while pressing and passes after releasing the button. The LMM outputs textual descriptions or answers, which are then converted into speech using the pyttsx3 text-to-speech engine~\cite{pyttsx3-doc} and played through the earphones worn by users, ensuring that only BLV participants can hear and interact with the LMM.

\section{Methods}
We conducted a mixed-methods study through four workshop sessions to explore how BLV and sighted participants use and perceive the smart glasses. In each session, two BLV and two sighted participants first take part in the workshop activity supported by \textit{CollabLens}. They then participate in a survey and a semi-structured interview separately. The total session lasts about two hours. The study protocol was reviewed and approved by the Institutional Review Board (IRB). 
\begin{figure*}[b]
    \centering
    \includegraphics[width=\textwidth]{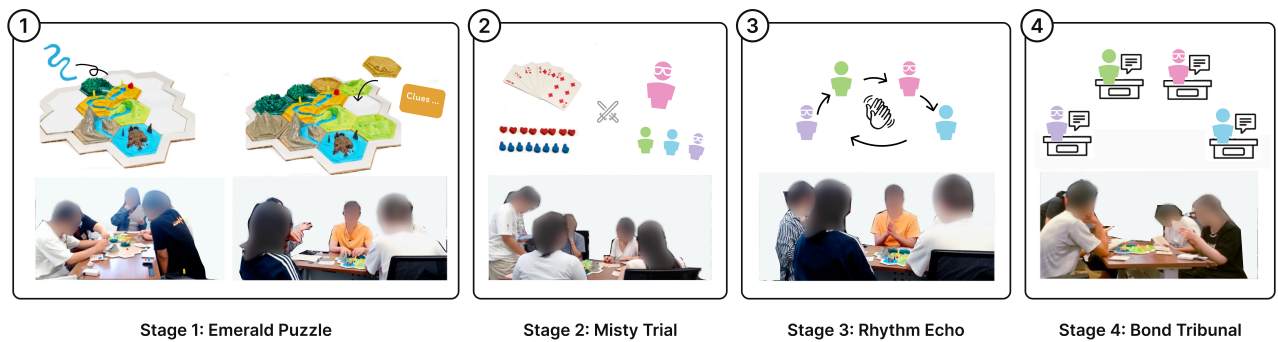}
    \caption{This figure shows the process of this workshop activity, including (1) Emerald Puzzle, (2) Misty Trial, (3) Rhythm Echo, and (4) Bound Tribunal. Each stage represents varying collaborative tasks that are common in mixed-vision activities.}
    \Description{This figure shows the process of this workshop activity, including four stages (1) Emerald Puzzle, (2) Misty Trial, (3) Rhythm Echo, and (4) Bound Tribunal. Each stage represents varying collaborative tasks that are common in mixed-vision activities.}
    \label{fig:Game}
\end{figure*}

\subsection{Participants}
We recruited 16 participants (8 BLV and 8 sighted people) through a long-term partnership of a Chinese college with our research lab and word-of-mouth.
Participants were aged 18-30 ($M = 24.75$, $SD = 3.46$), with 11 identified as female and 5 identified as male. They were all students pursuing an undergraduate or graduate degree and fluent in Mandarin. 
All eight BLV participants were completely blind with no usable vision. They had prior experience using screen reader and AI-powered video chat applications, and reported basic understanding and familiarity with smart glasses. All participants reported above average comfort level with mixed-vision collaboration. We detailed their prior experience of mixed-vision collaboration in Table~\ref{tbl:participant_details}.


\begin{table*}[t]
\centering
\small

\begin{tabularx}{\textwidth}{c c C C}
   \toprule
    Session & Participant ID & Self-reported visual ability & Prior Experience of Mixed-Vision Collaboration\\
    \midrule
    1 & 1A & Lost vision during adolescence; retains light perception without shape discrimination. & Have experience in daily affairs.\\
      & 1B & Lost vision during adolescence; retains light perception without shape discrimination. & Have experience in learning contexts.\\
      & 1C & Sighted & Extensive experience in daily affairs, professional, and learning contexts.\\
      & 1D & Sighted & Limited experience in professional contexts.\\
    \midrule
    2 & 2A & Lost vision during childhood; no light perception. & Have experience in daily affairs, professional, learning contexts and social activities.\\
      & 2B & Blind since birth; retains limited residual vision for large or blurred shapes. & Have experience in social activities.\\
      & 2C & Sighted & Limited experience in daily affairs.\\
      & 2D & Sighted & No prior experience.\\
    \midrule
    3 & 3A & Lost vision during childhood; retains light perception without shape discrimination. & Have experience in daily affairs, professional and social activities.\\
      & 3B & Lost vision in infancy or early childhood; no light perception. & Have experience in daily affairs, professional, learning contexts and social activities.\\
      & 3C & Sighted & No prior experience.\\
      & 3D & Sighted & Limited experience in professional contexts.\\
    \midrule
    4 & 4A & Lost vision during adolescence; retains limited residual vision for large or blurred shapes. & Have experience in professional and social activities.\\
      & 4B & Blind since birth; retains light perception without shape discrimination. & Have experience in daily affairs and social activities.\\
      & 4C & Sighted & Extensive experience in learning, daily affairs, and professional contexts.\\
      & 4D & Sighted & No prior experience.\\
    \bottomrule 
\end{tabularx}
\caption{BLV Participants' Demographic Details by Session}
\label{tbl:participant_details}
\end{table*}

We divided them into four groups based on their availability, with each group consisting of two BLV participants and two sighted participants. 
As we recruited participants through a long-term collaboration with a local institute, most participants knew at least one member in the group prior to the workshop. Each group will participate in the workshop session together. All participants were informed of the study details and provided consent. Additionally, sighted participants receive ethical training before participating in the study. Sighted and BLV participants receive compensation of 100 CNY and 200 CNY, respectively.

\subsection{Workshop Activity}

\label{workshop}


Our workshop design goal is to simulate and elicit the full range of challenges in mixed-vision social interactions while remaining engaging and naturalistic. To this end, we selected tabletop games as the study context. These games require participants to manipulate diverse tangible tokens (e.g., boards, cards, tiles, and pawns) and involve varying modes of interpersonal interaction (e.g., social deduction, collaborative decision making), making them an ideal environment for examining flexible and complex social dynamics \cite{bolesnikov2022understanding, thevin_inclusive_2021,lasley2025dungeons,pickova2022games}.
Meanwhile, the engaging and enjoyable nature of gaming can reduce the laboratory effect where participants could behave differently when conscious of being studied, providing more valid empirical insights about technology use in social settings \cite{allen2024using}.

We created our workshop activity based on tabletop games, integrating existing mechanisms and elements to construct a representative setting for mixed-vision social activities (for the whole game flow, see supplementary materials). 
We based the game narrative on the well-known story ``The Wizard of Oz" \cite{Baum_2008} and curated the game \textit{Echoes of Chaos}. 
The game included four roles: Dorothy, Scarecrow, Tin Man, and Cowardly Lion. Each role have varying attributes in Courage and Wisdom which can influence the overall storyline and final outcome. 
In each session, the four participants would each draw a random role and collaboratively complete the tasks in the four stages of the game (Figure~\ref{fig:Game}): 
  
\begin{enumerate}
  \item \textit{Emerald Puzzle}. Players use tactile clues and hint cards to assemble 3D tiles into the map of Emerald Land.
  \item \textit{Misty Trial}. Players need to draw a card and add its value with either their Courage or Wisdom attributes to pass the challenge. The outcome will influence the task difficulty of following chapters.
  \item \textit{Rhythm Echo}. Players take part in a rhythm relay challenge, repeating and adding rhythm patterns in sequence within a limited number of rounds to complete the task.
  \item \textit{Bond Tribunal}. Players engage in a role-based debate and vote, collectively determine the final outcome.
\end{enumerate}

See supplementary materials for the detailed facilitation scripts within each stage of the game.

\subsection{Apparatus and Procedure}

\begin{figure}[h]
    \centering
    \includegraphics[width=\linewidth]{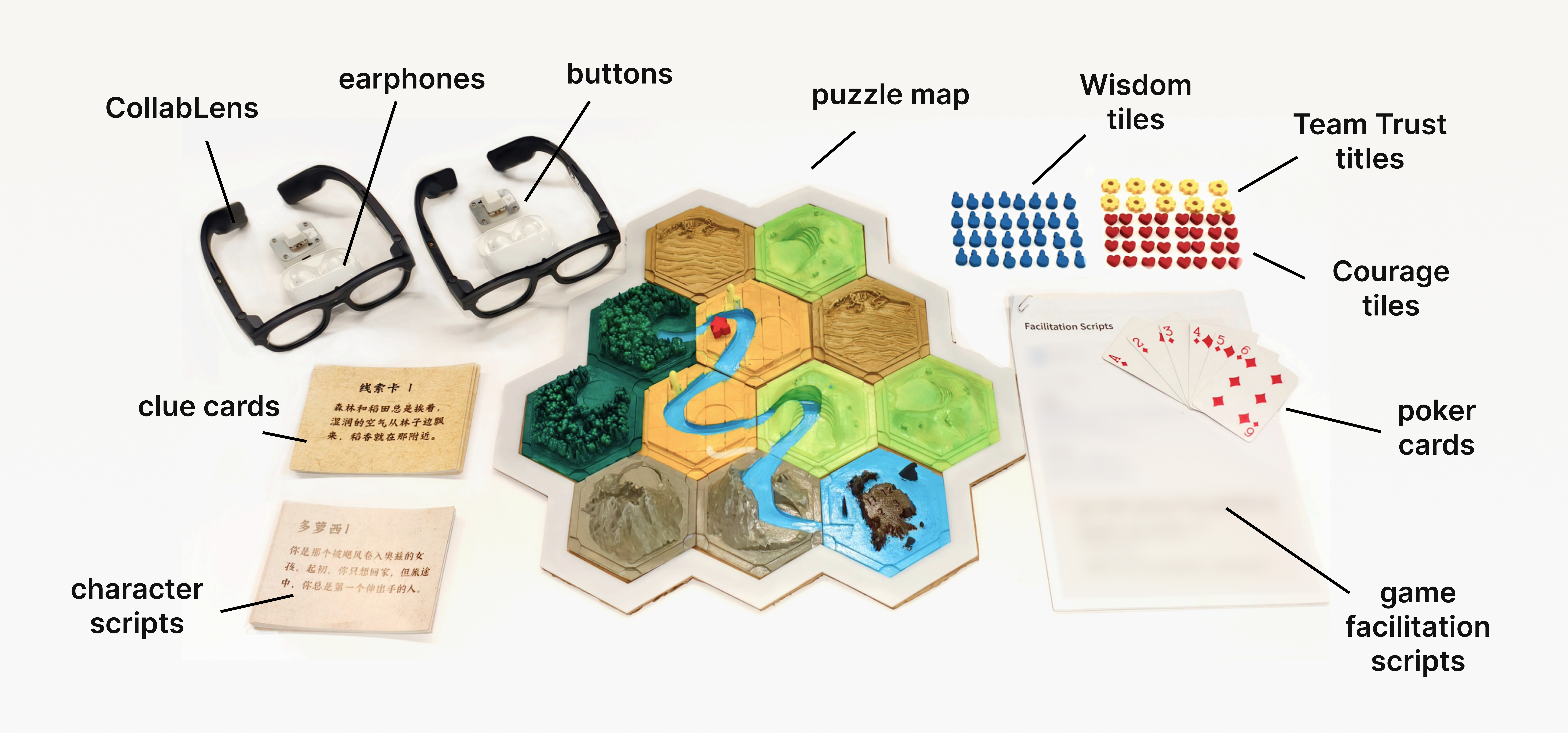}
    \caption{The key materials utilized in the session include Technical Prototype (smart glasses, earphones, button); Game Materials (character scripts, clue cards, puzzle map, wisdom points, team trust points, courage points, card check), and Facilitation Tools (facilitation scripts for facilitator). }
    \Description{The key materials utilized in the session includes Technical Prototype (smart glasses, earphones, button); Game Materials (character scripts, clue cards, puzzle map, wisdom points, team trust points, courage points, card check) and Facilitation Tools (facilitation scripts for facilitator).}
    \label{fig:Apparatus}
\end{figure}

For each session, we have one facilitator from our research team as the host for the workshop session. 
We prepared smart glasses prototypes and workshop materials in advance for each session (See Figure~\ref{fig:Apparatus}), and followed procedure protocols to ensure data consistency. The total session lasts about two hours, and it was divided as follows:

\textbf{Introduction (20 minutes).} Upon participants' arrival, we would briefly introduce the study and confirm oral and written consents from participants. The participants would first complete a short demographic survey. The facilitator then presented \textit{CollabLens} and guided BLV participants to experience the functionalities detailed in Section~\ref{system design}.
In this stage, we would answer any clarifying questions from participants to make sure they clearly understand how to use the device. We would also note that there is no prescribed way of using the device and they should 
The participants are encouraged to introduce themselves to each other to get familiar with the research group.

\textbf{Workshop session (70 minutes).} Centering around the tabletop game activity (Section \ref{workshop}), the participants would randomly draw a character and collaboratively finish the game tasks with the support of \textit{CollabLens}. 
We collected observational notes for data analysis later.
In this stage, we adopted a non-intrusive approach. We would not provide further guidance on how they could specifically use \textit{CollabLens} to finish the game tasks to avoid intervening in the participants' interaction dynamics. This allowed us to observe how participants naturally adopt and interact with \textit{CollabLens} and uncover real-world use patterns rather than prescribed ones.

\textbf{Survey (10 minutes).}
We administered the survey using the Interpersonal Regulation Interaction Scale (IRIS) \cite{swerdlow2022interpersonal} and the Perceived Group Inclusion Scale (PGIS) \cite{jansen2014inclusion} after the workshop session. We used IRIS to evaluate participants’ perceptions of how smart glasses supported their interpersonal interactions, and employed PGIS to assess their sense of group inclusion during the workshop. While we used the same PGIS questionnaire for both groups, we slightly adjusted the phrasing of IRIS questions for BLV and sighted participants respectively. For example, we asked BLV participants how they felt when using the smart glasses, whereas we asked sighted participants how they felt when their BLV peers were using the smart glasses.

\textbf{Semi-structured interview (30 minutes).} After completing the workshop, the participants would take part in a one-on-one interview separately. Due to the exploratory nature of our study, we adopted an semi-structured interview approach. These interview explored participants’ perceptions of \textit{CollabLens}, their collaborative experience, and their reflections on how smart glasses could mediate accessibility and social interaction. 
Our one-on-one interviews were conducted in Mandarin, and each interview was audio recorded with participants’ consent. Each interview session lasts around 25 minutes.
All sessions were recorded and transcribed for later qualitative analysis.

\subsection{Data Collection and Analysis}

We collected a mix of data across all study phases, including audio and video recordings, semi-structured interviews, observational notes, and a survey. Two researchers from our team observed the workshop activities while taking detailed notes, focusing specifically on how BLV participants interacted with the smart glasses and their sighted peers. The facilitator for each interview was responsible for organizing the transcripts. 
We then translated the observation notes and interview transcripts into English using ChatGPT-5, and the first two authors manually reviewed all transcripts for accuracy. We applied thematic analysis to those qualitative data \cite{braun2012thematic}. Two authors independently generated initial codes from the original transcripts through open coding, followed by discussions to reach consensus on interpretations and to group similar codes into clusters. Emerging themes were identified based on internal connections, and final themes were collaboratively refined using an affinity mapping approach \cite{olson_curiosity_2014}.

For the quantitative analysis, we conducted descriptive and inferential statistical analyses to compare responses between sighted and BLV participants on their group inclusive experiences during our workshop activities. Referring to the original scale, our PGIS results were analyzed in four dimensions, \textit{Group Membership}, \textit{Group Affection}, \textit{Room for Authenticity}, and \textit{Value in Authenticity}, using a 5-point scale (1 = Strongly Disagree; 5 = Strongly Agree) and our IRIS results were analyzed in four dimensions, Responsiveness, Hostility, Cognitive Support, and Physical Presence, using a 9-point scale (1 = Strongly Disagree; 9 = Strongly Agree). 
For each subscale, we computed the mean ($M$) and standard deviation ($SD$) for both groups. To examine potential group differences, we conducted independent-samples $t$-tests on each subscale score and calculated Cohen’s $d$ to estimate the effect size of each comparison. We did not intentionally seeking statistical significance, but we used these results to illustrate how smart glasses supported inclusive experiences and their interpersonal interactions. 
\section{Findings}

Our quantitative data first indicates that smart glasses broadly supported inclusive and positive interpersonal experiences for both groups (Section~\ref{Quant Finding}). These quantitative data complement the qualitative findings and reduce potential subjective bias.
Our qualitative findings next present how BLV participants engaged with smart glasses during the workshop activities (RQ1), revealing that independent task completion dominated their usage patterns over collaborative social scenarios (Section~\ref{Scenarios}). 
We then examine the challenges and opportunities that emerged in their interaction experiences (RQ2) across two dimensions: the wearable affordances (Section~\ref{Affordance}) and the human-AI communication loop (Section~\ref{Interaction}).
Finally, we explore how both BLV and sighted participants perceived smart glasses' role in mediating interpersonal dynamics and facilitating inclusion in mixed-vision social contexts (RQ3, Section~\ref{perspectives}).

\subsection{Survey Results}
\label{Quant Finding}


Our quantitative data analysis results show that both BLV and sighted participants had relatively consistent evaluations and there were no significant differences ($p> .05$). 
The PGIS results indicate that both BLV and sighted participants reported comparably high scores across all four dimensions, with means generally exceeding 4.2 out of 5. Specifically, for \textit{Group Membership}, sighted participants scored $M = 4.59$ ($SD = 0.79$), while BLV participants scored $M = 4.53$ ($SD = 0.53$). \textit{Group Affection} yielded identical means of $M = 4.22$ for both groups. The \textit{Room for Authenticity} dimension also showed nearly equivalent scores (sighted: $M = 4.47$, $SD = 0.53$; BLV: $M = 4.50$, $SD = 0.52$), and \textit{Value in Authenticity} also demonstrated identical means of $M = 4.53$ (sighted: $SD = 0.71$; BLV: $SD = 0.49$). The consistently low standard deviations (all below 0.87) across both groups suggest stable and reliable perceptions of inclusion. 
These findings indicate that smart glasses facilitate inclusive collaborative experiences, wherein differences in visual ability do not disadvantage BLV participants in terms of group belonging or self-worth recognition.

\begin{table*}[ht]
\centering
\small
\caption{PGIS and IRIS Subscale Scores for Sighted and BLV Participants}
\label{tab:pgis_iris_comparison}
\begin{tabular}{lcccccccc}
\toprule
\textbf{Subscale} & \multicolumn{3}{c}{\textbf{Sighted ($n=8$)}} & \multicolumn{3}{c}{\textbf{BLV ($n=8$)}} & \textbf{$t$} & \textbf{Cohen's $d$} \\
\cmidrule(lr){2-4} \cmidrule(lr){5-7}
 & $M$ & $SD$ & & $M$ & $SD$ & & & \\
\midrule
\textit{PGIS} & & & & & & & & \\
\quad Group Membership & 4.59 & 0.79 & & 4.53 & 0.53 & & 0.186 & 0.093 \\
\quad Group Affection & 4.22 & 0.81 & & 4.22 & 0.63 & & 0.000 & 0.000 \\
\quad Room for Authenticity & 4.47 & 0.53 & & 4.50 & 0.52 & & -0.120 & -0.060 \\
\quad Value in Authenticity & 4.53 & 0.71 & & 4.53 & 0.49 & & 0.000 & 0.000 \\
\midrule
\textit{IRIS} & & & & & & & & \\
\quad Responsiveness & 7.48 & 1.16 & & 7.56 & 1.16 & & -0.137 & -0.068 \\
\quad Hostility & 1.43 & 0.85 & & 1.30 & 0.48 & & 0.362 & 0.181 \\
\quad Cognitive Support & 6.73 & 2.47 & & 8.05 & 0.97 & & -1.410 & -0.705 \\
\quad Physical Presence & 7.13 & 1.55 & & 5.56 & 2.29 & & 1.597 & 0.799 \\
\bottomrule
\end{tabular}
\end{table*}

In terms of how participants perceived smart glasses as supporting their interpersonal social interaction, the IRIS results revealed more nuanced differences between BLV and sighted participants across the four subscales. For \textit{Cognitive Support}, BLV participants demonstrated slightly higher scores ($M = 8.05$, $SD = 0.97$) compared to sighted participants ($M = 6.73$, $SD = 2.47$), suggesting that BLV participants perceived greater problem-solving assistance from smart glasses during collaborative activities. Conversely, \textit{Hostility} scores remained minimal for both groups (sighted: $M = 1.43$, $SD = 0.85$; BLV: $M = 1.30$, $SD = 0.48$), with means well below 2 on the 9-point scale, indicating negligible negative interactions regardless of visual ability. \textit{Physical Presence} revealed an interesting pattern, with sighted participants reporting higher scores ($M = 7.13$, $SD = 1.55$) than BLV participants ($M = 5.56$, $SD = 2.29$). Finally, \textit{Responsiveness} showed nearly identical scores between both groups (sighted: $M = 7.48$, $SD = 1.16$; BLV: $M = 7.56$, $SD = 1.16$), suggesting that smart glasses effectively supported appropriate interpersonal exchanges. 
In general, these results highlight that smart glasses supported both inclusive group experiences and interpersonal interactions consistently across BLV and sighted participants.

\subsection{Engaging with Smart Glasses During Activities}
\label{Scenarios}
Across our observations and interviews, we first found that our BLV participants primarily used smart glasses for text reading, tangible object recognition, and environmental awareness 
. Notably, these tasks are typically completed independently by the users and do not involve interaction or cooperation with other participants.


Among the usage scenarios, \textbf{\textit{text reading}} was the most prominent and consistent; for example, they used it to read text on character cards and scripts. This emphasis may have been driven by the text-reading segments required for interpreting clues and scripts to keep the workshop progressing. Despite this, BLV participants also indicated in the interviews that text reading remained a significant scenario beyond the specific demands of the workshop activities. 

In addition, we observed that BLV participants attempted to use the smart glasses for \textbf{\textit{tangible object recognition}}. Although they could perceive the shape of objects through holding it, information such as color was not accessible through tactile exploration alone. To address this gap, they used the glasses to obtain descriptions. We also found that participants often posed open-ended questions in these situations, such as \textit{``What is this?'' } rather than making more attribute-specific requests like \textit{``Please describe the color of the object in my hand.''}


Moreover, \textbf{\textit{environmental awareness}} describes situations in which BLV participants used smart glasses to access information about the surrounding environment, including the arrangement of objects, the interior decoration, and the items or figures occupying the largest portion of the frame. 
These descriptions often prompted subsequent use of the object recognition feature. For instance, when the smart glasses reported, \textit{``This is an office scene with a hexagonal object on the desk, possibly used for gaming...''} Participants frequently followed up with a long press to ask, \textit{``What are the features of this hexagonal object?''}


Besides these usage scenarios, we also found that in the introduction stage, when BLV participants were instructed in using \textit{CollabLens}, they would prompt the glasses to describe other participants’ appearance, expressions, and movements, but such uses were not sustained when the workshop session officially started.
What's more, they also expressed limited interest in using smart glasses to interpret non-verbal social cues. For instance, Participant 3A ranked the three scenarios mentioned above as top 3, excluding non-verbal cue interpretion as they are "less frequent" in actual usage. 
Participant 1B explained: 
\begin{quote}
\textit{``...when we are with sighted friends, we don't always need to know their facial expressions, cause if the person [you are talking to] is truly happy, angry, or sad, we can tell from the tone of their voice...}
\end{quote}

\subsection{Challenges and Opportunities on Wearable Affordances}
\label{Affordance}


In general, we found BLV participants highlighted the convenience that smart glasses' wearable affordances brought. 
Participant 4A compared smart glasses with existing AT like screen readers, OCR apps, and other AI-based systems, and explained, \textit{``...[Smart] glasses are more convenient, because on a phone you have to tap, to take a photo, then tap again to recognize it... It [smart glass] is less cumbersome, yeah, more convenient.''} 
Sighted peers also echoed these sentiments. For example, Participant 3C acknowledged that when compared to phones, smart glasses ``...are definitely more convenient to wear, unlike a phone, which requires at least one hand to hold up... glasses worn over the eyes would be incredibly practical.'' 

Another dimension mentioned by the sighted participants was the unobtrusive appearance. They described the device as similar to everyday eyewear, thereby minimizing visible differentiation within the group and facilitating inclusive integration. 
For example, Participant 1C commented:

\begin{quote}
\textit{``I think eyewear design is a great idea... although those glasses look a bit bulky and have a cable, if they didn't have the cable or the frame were a bit thinner, they wouldn't look much different from ordinary glasses... I think they can bridge the gap between certain BLV groups and ordinary people.''}
\end{quote}

Participant 1D stated, \textit{`` ...These [smart glasses] aren't an unusual item; everyone wears them, so wearing them doesn't make them [BLV participants] stand out from the crowd.''}
Participant 3D also reflected on these aspects, \textit{``I feel the [smart] glasses carry a certain symbolic meaning, much like how blind people wear sunglasses nowadays... For the general public, I believe it [smart glasses] serves as a reminder and can spark societal concern.''} However, none of the BLV participants mentioned the appearance of smart glasses in strengthening their sense of inclusion.

Despite these positive aspects, we also found that hardware design limitations affected participants’ willingness to use the device and compromised its sustainability. The most prominent complaint concerned the device's weight and wearing comfort. BLV participants frequently described the smart glasses using terms such as ``too heavy,'' ``uncomfortable on the ears,'' and adds pressure on [them],'' which led to a desire to remove the device shortly after wearing it. 

Participants also expressed anxiety regarding the limited battery life, noting that some smart glasses would shut down unexpectedly due to low power. Participant 4A further mentioned the discomfort brought by the heat of the battery.

Additionally, the camera's position often confused users. Despite the first-person perspective, placing the camera on the forehead rather than aligning it with the line of sight made it difficult to accurately capture objects. For example, Participant 4B complained, \textit{``...Even if it's slightly crooked, tilted, or has reflections, it can be very difficult [to capture the image].''} We also observed that some BLV participants would lean their ears closer to objects they wished to have the glasses recognize, likely influenced by prior use of screen readers. This posture did not align well with the camera’s perspective, adding friction to the overall user experience.

\subsection{Challenges and Opportunities Across the Communication Loop}
\label{Interaction}

In the interaction flows of BLV participants and smart glasses, we observed a number of challenges and opportunities for better experience. We mapped them to the different stages of information transmission: user's input, LMM generation, audio delivery, and user feedback. The interactions in these stages shaped together how seamlessly smart glasses could be integrated into mixed-vision social activities.

\subsubsection{Potential Social Burden for Voice Input}

Participants reported that relying on voice input in shared spaces could introduce social pressure. BLV participants worried about drawing unwanted attention, slowing others down, or having to repeat themselves in public. Participant 1C observed, ``...They are now always afraid of disturbing us, or that their voices might be too loud or somehow inappropriate...'' This embarrassment was intensified when the system failed to respond, forcing users either to speak louder or to issue multiple commands in front of others.

Some BLV participants felt that addressing the glasses instead of directly engaging with people around them was inherently unnatural. Participant 3B noted:
\begin{quote}
\textit{``...[In a social setting] asking the [smart] glasses 'What are they doing right now?' would be less natural than just asking the person 'What are you doing right now?'... If I'm mid-conversation and suddenly stop and ask the [smart] glasses, 'What is she doing right now? Is she trying to poison my drink?' [laughing] It just feels weird.''}
\end{quote}
As a result, several participants expressed a preference to reserve smart glasses for private or low-stakes situations, and to fall back on familiar practices (e.g., asking sighted peers or using their phone) when social stakes were high.

To mitigate these issues, participants envisioned more discreet and controllable input modalities. For example, Participant 1B proposed physical controls that better support turn-taking, such as a press-to-speak mechanism: \textit{``...So I think we could add a feature where you press once to start speaking, then press again to pause... otherwise the interaction gets too chaotic...''} Meanwhile, Participant 3C envisioned the potential of gesture-based triggers: \textit{``...If I point my finger like this, it starts reading... This would be more socially unobtrusive than repeatedly giving commands in public.''} These suggestions highlight a desire to shift from always-on or repeatedly voiced commands toward input methods that are both socially acceptable and easy to coordinate with ongoing conversation.

\subsubsection{Reliability, Information Density, and Latency Issues of LMM Generation}

Participants frequently mentioned the reliability of recognition and reasoning as a critical concern. Participants reported that smart glasses occasionally produced contents that were only partially correct, irrelevant, or entirely fabricated. Participant 2B commented,
\begin{quote}
\textit{...(AI) read half of it correctly, then might not see the next half clearly and start making up things based on the first half. I do not demand word-for-word accuracy, but don't just make things up either.''}
\end{quote} 
In one instance, when Participant 4A asked the glasses to read a character card, the system misaligned the visual input and generated a fabricated description of a non-existent character “Xiao Ming,” complete with a birthday and favorite color.
We observed that such errors directly undermined user trust and caused hesitation in relying on the system, especially when users could not easily verify the output themselves. Participant 2D reflected: 
\begin{quote}
\textit{I think this might put them in an embarrassing situation... The smart glasses gave them incorrect information, and then when we provided the correct information, I think they might express some frustration or something similar in their verbal responses...''}
\end{quote}

Response latency emerged as another primary challenge, which could disrupt the natural rhythm of activities, often leaving awkward pauses in conversation. Participants 2A and 3B expressed worries that they were slowing down the group. Participant 2B described, \textit{``...waiting half a minute to reply [...] indeed leaves others hanging for too long...''} Participant 1A explained, \textit{``...as it has very high real-time requirements, the shorter the delay time, the better...''} 
Participants also criticized the verbosity and delivery pace of the system's descriptions, which further prolonged their waiting time. 
Participant 3A felt descriptions contained unnecessary or overly long phrasing: ``...for example... if it [says] something is green, it will then go on to say it's like a lawn, or something else... with a lot of unnecessary details. It's just a lot of rambling.'' Participants 1B, 3A and 4A suggested higher speech rate. Over half of BLV participants compared \textit{CollabLens}' response style to Doubao, a mobile AI chatbot that enabled video chat \\\footnote{AIbase. \textit{DouBao Launches Video Call Function: Supports Real-time Interaction and Integrates Visual and Language Input Capabilities.} Retrieved January 18, 2026 from 
https://news.aibase.com/news/18340.}
. Participant 1B detailed: 
\begin{quote}
``...The previous response could be made softer and more like natural human speech... it should answer directly after I ask a question... you could completely refer to Doubao as an example. I think Doubao is excellent because it answers exactly what you ask and doesn't include any unnecessary words...''
\end{quote}

\subsubsection{Navigating Social Soundscapes for audio deliveries}

Participants shared challenges regarding audio delivery. Some participants felt unable to hear the surrounding environment with earphones on. Participant 2A described:
\begin{quote}
\textit{“...I feel there's no need to make it into earphones; you could just integrate a small speaker into the camera module... Because with these noise-canceling headphones, I can't hear anything at all.  It's like we're blindfolded and have our ears covered at the same time.”}
\end{quote}
In sessions 3 and 4, the discomfort led participants to take off the earphones and use the default speakers on smart glasses instead. Yet new concerns about the timing and volume of audio output arose. that some BLV peers lowered their voices when prompting the glasses to avoid disrupting others. As Participant 4D observed, \textit{``...since they need to play it out loud, they tend to avoid [using the glasses too frequently]...''}

\subsubsection{User Feedback and Strategy Repair}
 
When facing conversational breakdowns with the smart glasses (e.g., no response, delayed response, or irrelevant answers), BLV participants often re-queried the device multiple times in an effort to obtain a satisfactory result by themselves. Participant 1A described this process as a matter of luck. They also mentioned their cognitive load that might be another factor for delay. Half of the BLV participants mentioned they needed to listen multiple times to remember the information. Participant 4A explained: 
\textit{``...Sometimes I might not catch everything the first time, and sometimes I need to have it read a few times. Yeah, sometimes after it finishes reading, I still don't get it...''}

When confidence in the technology declined, a common response was to revert to pre-existing support channels. For example, participant 2B sought direct assistance from sighted peers after experiencing repeated failures. Participants 2A, 2B, and 3A frequently asked peers, ``Is what it said correct?'' to verify information and reduce the risk of error. Sighted peers also noted this behavior, commenting on the increased reliance on them for confirmation. For instance, Participant 2C observed: 
\begin{quote}
\textit{``...When they [BLV participants] experienced a delay or felt that the situation remained unclear even after waiting, or when the smart glasses failed to provide effective game instructions, they would naturally express the need for human intervention, as observed in our study, where participants explicitly requested, `Could someone help read this for us?'...''}
\end{quote}

\subsection{Perceiving Smart Glasses' Role in Mixed-Vision Social Contexts}
\label{perspectives}

\subsubsection{Smart Glasses' Role in Mediating Interpersonal Dynamics}
BLV participants appreciated smart glasses in bridging the information gap and reducing their reliance on other member's help. Participant 1A described the purpose of smart glasses in mixed-vision collaboration as ``help[ing] us move from not being able to do something to being able to do it.''
They appreciated smart glasses in enabling them to flexibly initiate tasks on their own before deciding whether to involve other tools or people. 
Participant 1B commented, \textit{``...With these glasses, I no longer need to ask my sighted friends to help me read. I can just use these glasses to read myself....''}
Participant 4B also stated, \textit{“...I do not need to seek help from [my group members]...” }
Participant 3A mentioned the simultaneity where each of them working on their own tasks at the same time, while Participant 3B emphasized the symmetry of the information the mixed-vision group get. Participant 3B further explained the nuance: \textit{``...It's like when you [and] your friend play, you always feel like you're asking them for a favor. We should play games 'together', right?...''} Participant 1B also described the greater agency they could gain: \textit{``...in the future, whether it's during meals, attending classes, studying together, or even going shopping, if I can obtain the information myself, I can then express [more of] my own thoughts or opinions...''}

While the technology provided blind users with visual information, the glasses introduced ambiguity in help-giving norms.
Several sighted participants reported feeling the natural helping dynamics were confounded. 
Participant 2D said, \textit{``... [watching BLV peers trying glasses] increases my own sense of not knowing what to do, because honestly, I don’t really know whether [the device] works or not...}'' They also expressed uncertainty about when and how to offer help: \textit{``...I generally won't interfere with [BLV peers'] device adjustment because I'm afraid helping will create more chaos...''}
Participant 2C noted that mediated help via the glasses left them ``sitting there doing nothing,'' and ``there's a lack of connection and a sense of inclusion,'' although they complemented their BLV peers should feel more inclusive.
In contrast, other participants expressed how smart glasses made them feel more connected. For example, Participant 3C described:
\begin{quote}
\textit{``...actually it's hard to know [BLV friends' side] of the world, and neither [can they], but with the help of these glasses, maybe we can let our worlds be shared with each other...''} 
\end{quote}

Participant 1D also described:
\begin{quote}
\textit{``...[Smart glasses] in a way built a bridge for us, enabling us to sometimes play games together. And throughout this process, it actually helped me get to know these BLV friends a little better...''}
\end{quote}


\subsubsection{Smart Glasses' Role in Facilitating Inclusion}

Participants generally view smart glasses as as assistive helpers that supplement and enhance their capabilities. They used expressions like ``little helper'' (1B) ``assistant'' (1A, 1B, 4C) and ``secretary'' (4C) and ``auxiliary'' (3A, 4B, 4D) to describe the glasses' supportive role. Several participants emphasized the glasses as intermediaries or bridges. Participant 3D described them as ``translator'' while Participant 4A described them as ``something like a tour guide,'' both highlighting smart glasses' role in interpreting the world for them.

We observed a tendency where sighted participants generally expressed more positive views regarding the smart glasses’ contribution, whereas BLV participants’ opinions were often more reserved. 
Sighted participants generally believed that the technology would help BLV participants engage more comfortably and confidently by satisfying their needs more flexibly through the device. For example, regarding expression, Participant 3D reflected, \textit{``...[the glass] seemed to help them more... although we tried hard to help them with tactile exploration... we can’t fully put ourselves in their shoes. So it feels like with these glasses, they could express themselves directly. ''} Participant 4D similarly highlighted the link between independent access and social engagement, stating, \textit{“They could now obtain information on their own that was previously inaccessible, and through that, express themselves, which might also make them more willing to communicate with others.”} Visually, participants also noted that the glasses’ everyday appearance helped normalize their use, thereby minimizing visible differentiation within the group (see Section~\ref{Affordance}).

In contrast, BLV participants expressed more reserved perspectives. While they acknowledged the practical value of smart glasses for supporting convenient information access, they less frequently framed the technology as altering their sense of social inclusion. 
When asked about expectations for future capabilities, BLV participants felt that even if such features were fully developed, their frequency of use and impact on inclusion would remain limited. One key reason was that they already developed alternative strategies for communicating with sighted peers through experience. For example, Participant 3A said, \textit{"...Unless it [smart glasses] functions better than [AI video chat apps] on phones, it's OK with phones only."} BLV participants perceived inclusive experiences as emerging more from interpersonal adaptation and mutual awareness than from technological mediation by smart glasses.
For instance, Participant 1B believed there is no currently necessity of applying smart glasses on social integration:
\begin{quote}
   \textit{``...if you go out with a sighted friend, they won't deliberately use gestures or body language. Of course, when meeting someone new or just getting to know them, they might instinctively use such cues. But once they realize you're not particularly sensitive to these signals, they'll naturally adjust their communication style to verbal or other understandable ways...''}
\end{quote}

\section{Discussion}

In this section, we first examine how BLV participants' preference for independent task completion reflects the importance of enriching interdependent assistive networks (Section~\ref{enrich}). We then analyze the challenges and opportunities for achieving seamless interaction in social contexts, considering hardware affordances, AI reliability, and interaction modalities that minimize social disruption (Section~\ref{envision}). Next, we explore smart glasses' role in creating reciprocal inclusion dynamics of mixed-vision collaboration, affecting not only BLV users but also sighted peers' roles and experiences (Section~\ref{enhance}). Finally, we discuss limitations of current research and future directions (Section~\ref{dis:limitations}).

\subsection{Enriching Assistive Network Beyond Independence}
\label{enrich}
We observed that BLV individuals preferred using smart glasses in individual scenarios (Section~\ref{Scenarios}) and completing tasks on their own without having to constantly require help from sighted peers (Section~\ref{perspectives}). Their emphasis on more independent task completion accounts for prior work's dominant focus on individual task scenarios \cite{li_oscar_2025, zhang_enhancing_2025, xia_ibgs_2022, li_more_2025}. Our study extends this understanding by situating these preferences within mixed-vision social contexts, revealing that the drive for doing things on their own still dominants when collaboration is the explicit goal.
This emphasis on independence appears rooted in BLV participants' concerns about burdening others and experiencing social pressures. This resonates with literature on help-seeking stigma among people with disabilities \cite{Shinohara_Wobbrock_2011, 10.1145/2700648.2809864}, where requesting assistance, even in collaborative environments, can generate psychological costs including feelings of dependence, vulnerability, and social debt. 

Although prior work discussed the important challenges in interpreting non-verbal social cues \cite{Buimer_2017, Qiu_Hu_Han_Osawa_Rauterberg_2020, Shinohara_Wobbrock_2011}, BLV individuals revealed less expectation in such application scenario compared to individual task completion, even if in social contexts \cite{Ruffieux_Hwang_Junod_Caldara_Lalanne_Ruffieux_2023}. 
As they have developed sophisticated alternative strategies for social communication, pursuing an absolute symmetry of information BLV and sighted participants may risk privilege visual norms and fail to address higher priority needs \cite{teng_help_2024}. Long term field studies could contribute to exploring how their strategies evolve as smart glasses become embedded in their daily assistive networks. 

Similarly, we noted the transfer challenge where established accessibility strategies may conflict with the design assumptions that are rooted in sighted experiences \cite{Turkstra_2025, Botelho_2021}. For example, some BLV participants leaned their ears toward sound sources while interacting, which might misalign the intended field of view and compromised the hands-free affordances. Although the forehead-mounted camera works well in outdoors scenario \cite{Lee_Sato_Asakawa_Asakawa_Kacorri_2021}, our indoor collaborative scenarios involve more complex elements of interest and require more precise attribution of attention.
Thus, designing for alignment with users’ embodied practices is essential for both accessibility and adoption. 

Intriguingly, although BLV participants appreciated the contribution of smart glasses, they saw inclusion as something rooted more in interpersonal dynamics rather than technological capabilities, and valued more on sighted peers' friendliness and adaptability in their inclusive experiences (see Section~\ref{perspectives}). While they hoped to reduce reliance on help from sighted peers, they did not view smart glasses as substitutes for human support or a way to help them accomplish tasks entirely independently. Instead, they viewed smart glasses as opportunities for a convenient addition to their existing assistive networks. 
This resonates with prior accessibility technology research discussing independence and interdependence
\cite{Bennett_Brady_Branham_2018, 10.1145/2700648.2809864, Hamraie_Fritsch_2019}. The interdependence framework reveals how smart glasses reconfigure the situated and simultaneous relationships between BLV individuals, sighted peers, and the environment, and challenge ability-based hierarchies that position BLV individuals primarily as recipients of assistance.

Our work reveals implications for future researchers should aim for in mixed-vision contexts. Although independence is more emphasized in both prior work on smart glasses and BLV participants in this study, seeing the technology within the support networks is important. This does not mean advocating interdependence over independence \cite{Bennett_Brady_Branham_2018, Kelly_2013}, but focuses on assessing whether technology expands users' choice and flexibility within their assistive networks. This requires considering how the newly introduced technology fits into the existing network that interrelates mixed-vision individuals, available technologies and the environment.

A productive design direction is thus advancing smart glassed toward adaptive systems dynamically balancing independence support and social coordination, such as detecting when a user is mid-conversation and suppressing verbose output in favor of brief haptic acknowledgment. The context-sensitive adjustment poses design challenges around intervention intensity, timing, and privacy that will require participatory co-design with both BLV and sighted stakeholders.


\subsection{Envisioning Seamless Interaction in Social Contexts}
\label{envision}

We analyzed the challenges and opportunities towards seamless interaction in mixed-vision social contexts based on the affordances of smart glasses (Section~\ref{Affordance}) and the interaction flow (Section~\ref{Interaction}). 
Participants acknowledged the hands-free advantage of smart glasses for visual assistance, aligning with previous literature highlighting the benefits of wearable form factors for BLV users \cite{lee_interaction_2018, waisberg_meta_2024, katakwar_sight_2025}. However, they also mentioned that the smart glasses have not yet achieved the necessary level in weight, comfort, and battery life for everyday use \cite{Danielsson_Holm_Syberfeldt_2020,Larbaigt_Lemercier_2023, montuwy2019using}. For smart glasses to achieve widespread adoption in real-world contexts and bring actual change, manufacturers need to prioritize these ergonomic factors with the same rigor applied to computational capabilities. 

In terms of LMM capabilities, participants raised concerns of system reliability, where the recognition errors and fabricated content eroded trust and reduced their willingness to use. As we employed \textit{CollabLens} as a technology probe and did not focus on improving LMM performance in this study, these commonly acknowledged hallucination problems were within our expectations. However, such frictions took on heightened significance in social contexts.
Compared to conversational breakdowns in individual use contexts with VAs \cite{porcheron_voice_2018, luger_like_2016, bentley_understanding_2018}, errors in co-present interactions carry immediate visibility and social costs \cite{deterding_embarrassing_2015}. When the smart glasses misread a character card or fabricated a description, BLV participants not just failed to get accurate information but also experienced embarrassment, anxiety about slowing others down, and erosion of trust in the device.


Participants repeatedly emphasized social pressure as a central concern shaping their interaction with smart glasses. 
Voice input, audio delivery, response latency, and speech rate all introduced potential disruption to the natural flow of mixed-vision activities (see Section~\ref{Interaction}).
Unlike speaking to a person, addressing a device mid-conversation violates conversational turn-taking norms.
What's more, in controlled environments like our workshop, participants had pre-established rapport and awareness of the technology, which likely attenuated embarrassment. In uncontrolled, everyday settings like a café, classroom, or public transport, the social cost would plausibly be higher, with strangers interpreting device-directed speech as disruptive or peculiar.

As participants generally perceive smart glasses as supplementary assistant, interpersonal communication remains primary, requiring unobtrusive input and output of the device. \textit{CollabLens} used a combined video-and-audio pipeline, leaving other modality configurations unexplored. While audio-primary modes optimize feedback delivery latency, richer multimodal interaction such as gesture-based triggers, gaze-based activation, brief tonal cues, or haptic feedback might be more socially agreeable, \cite{kim2011designing,chang2024}.
Future work can have controlled comparisons of these configurations to yield actionable guidance for system designers.



\subsection{Enhancing Reciprocal Mixed-Vision Inclusion}
\label{enhance}
Smart glasses emerged as a new actor in established social dynamics, creating challenges for sighted participants to navigate uncertainty and adapt their role in collaboration. This uncertainty was compounded by the asymmetric information transmission, as sighted peers could not directly know what the smart glasses communicated to BLV participants. They were thus uncertain about whether the information conveyed was accurate, and how to calibrate their own assistance accordingly. BLV users become the information relay point, a responsibility that can be both empowering and burdensome. Prior works discussed shifting the educational burden from BLV users to sighted users and designing to help sighted users learn how to collaborate in help-centered contexts \cite{teng_help_2024}. Extending this to broad social interaction scenarios, we advocate taking the interaction between sighted users and smart glasses into consideration and enriching the channels through which BLV and sighted peers exchange information. 

The tension between form and function further shaped participants’ perceptions. While sighted peers viewed the glasses’ resemblance to conventional eyewear as a way to reduce stigma and enhance social acceptance, BLV participants prioritized functionality over appearance (Section~\ref{Affordance}). 
This divergence challenges the assumption that discreet or inconspicuous designs automatically promote inclusion and the need to balance aesthetics with accessible affordances \cite{DosSantos_2022,Culham_Nind_2003, Low_2020, Moser_2000}.
This finding supports recent calls to move from user-centered design toward proactive co-creation, where users directly contribute to shaping assistive technologies that fit their lived experiences \cite{Moon_Baker_Goughnour_2019}.

The normalizing appearance and unease around always-on cameras also creates tension. Prior research has documented BLV users' privacy concerns about assistive cameras recording their surroundings without their awareness and the social acceptability challenges of always-on cameras that may capture bystanders without notification \cite{ahmed2018up,akter2020privacy,profita2016effect,iqbal2023adopting}. Notably, no participants in our study raised these concerns. Instead, when discussing camera recording on smart glasses, some sighted participants reported feeling more connected to BLV peers as smart glasses' lens creates a bridge for them within the visual field. 

This contrast likely reflects two contextual factors. First, our study occurred in a controlled setting with a small group of participants who collaborated closely. As they participated with prior notification about the study scope, they had expectations and explicit awareness that smart glasses with cameras were in use. 
Second, as prior works suggested that cultural backgrounds play a role in varying attitudes toward privacy concerns of video camera-based technologies \cite{kim2025cultural}, we hypothesize participants' comfort with being recorded in this collaborative setting might be related to the cultural norms, though more research is needed to validate this hypothesis. 

Nonetheless, this absence of concern should not be read as evidence that the tension is resolved; rather, it underscores the importance of designing explicit camera-state signaling (e.g., visible indicator lights, privacy modes) and consent affordances before deployment in open social contexts where bystanders are not pre-briefed. Real-world application of smart glasses would not be naturally confined to specific scenarios. Therefore, even if designing features for a restricted task, designers should make sure the privacy considerations are applicable to all scenarios and venues.

\subsection{Limitations and Future Directions}
\label{dis:limitations}

Our study has several limitations. 
First, as \textit{CollabLens} adpoted a combined video-and-audio pipeline, it remains an open question how different modality configurations affect both usability and social dynamics. Exploring and comparing more interaction modalities, such as gaze-based or gesture-based alternatives, would provide actionale design guidance toward a seamless interaction experience. 

Second, our workshop activities occurred in structured, time-limited contexts designed to simulate mixed-vision social interactions, but they are not equivalent to everyday social life. The tabletop game format, while ecologically motivated, constrained the range of social dynamics that could emerge. Interactions were goal-directed and temporally bounded in ways that routine social encounters are not. It cannot fully capture how smart glasses mediate social norms over uncontrolled daily settings or extended use. 
Longitudinal field deployments across diverse social contexts would be particularly valuable for understanding how BLV users' strategies for deploying the device evolve as it becomes embedded in their daily assistive networks, and whether sighted peers' helping behaviors shift as smart glasses move from novelty to familiar presence. 

Third, our sample was drawn from a single academic institution, consisted entirely of totally blind participants with no usable vision, and skewed toward individuals with prior experience using assistive technology and relatively high comfort with mixed-vision collaboration. This likely produced a more technologically optimistic group than a broader population sample would. 
Future work should include broader populations like people with low vision, participants skeptical of assistive technology, and those from varied cultural backgrounds and support contexts. Research across these groups would both test the generalizability of our design recommendations and surface needs and concerns that our current sample may have underrepresented.

Finally, our study did not examine the broader social and structural factors that mediate inclusive experiences, such as organizational norms, physical environment design, or disability policy contexts.
Future work might examine how smart glasses function in different cultural contexts where norms around disability, assistance, and independence vary, or how organizational policies and physical environments mediate the technology's impact on inclusion.
\section{conclusion}

In this study, we developed a technology probe \textit{CollabLens} and introduced it into our workshop sessions to understand how smart glasses can contribute meaningfully to reshape interpersonal dynamics and facilitate inclusive participation in mixed-vision social activities. 
Our mixed-methods analysis indicated that BLV participants valued smart glasses primarily for expanding their assistive networks with more flexible, independent access to visual information. 
Realizing this potential in real-world contexts requires addressing challenges in seamless interaction, improving long time wearing comfort, enhancing reliability and trust, and minimizing social pressure. 
We highlights that designing for mixed-vision contexts demands attention to facilitate a reciprocal collaboration. 
Together, our work contributes empirical insights and design directions for creating smart glasses towards more equitable, reciprocal, and genuinely inclusive mixed-vision social experiences.

\section*{GenAI Usage Disclosure}
We used ChatGPT-5 to translate user quotes and study materials into English for inclusion in this paper. We also used Gemini 3 nano banana to generate illustrative figures for visualization and storytelling purposes. The authors take full responsibility for the use of GenAI and the integrity of all reported results in this work.

\begin{acks}
    This research was supported by the Foundation of the Ministry of Education of China (Grant No. 23YJCZH092), and the Research Project of China Disabled Person's Federation - on assistive technology (Grant No. 2024CDPFAT-04).
\end{acks}
\bibliographystyle{ACM-Reference-Format}
\bibliography{main}

\appendix

\end{document}